\documentclass{sig-alternate}
\usepackage[vlined]{algorithm2e}
\usepackage{xcolor}

\toappear{
   \hrule \vspace{5pt}
   Some Journal
}

\SetKwRepeat{Do}{do}{while}%

\begin{document}

\title{Clustering case statements for indirect branch predictors}

\numberofauthors{3}

\author{
\alignauthor
Evandro Menezes\\
       \affaddr{Samsung Austin R\&D Center}\\
       \email{e.menezes@samsung.com}
\alignauthor
Sebastian Pop\\
       \affaddr{Samsung Austin R\&D Center}\\
       \email{s.pop@samsung.com}
\alignauthor
Aditya Kumar\\
       \affaddr{Samsung Austin R\&D Center}\\
       \email{aditya.k7@samsung.com}
}

\maketitle
\begin{abstract}
  We present an $O(nlogn)$ algorithm to compile a switch statement into jump
  tables. To generate jump tables that can be efficiently predicted by current
  hardware branch predictors, we added an upper bound on the number of entries
  for each table. This modification of the previously best known algorithm
  reduces the complexity from $O(n^2)$ to $O(nlogn)$.
\end{abstract}


\section{Introduction}
In super-pipelined processors, there is a significant lag between the beginning
of the execution of an instruction and when its result becomes available. This
trait also applies to conditional direct branches, however, waiting for the
result of the instructions affecting a conditional branch is wasteful.
Therefore, the processor makes an educated guess as to where the conditional
branch will lead to.  If the guess is wrong, there is a serious penalty.
Conditional direct branch prediction techniques aim at increasing the frequency
when the processor makes the right guesses.

Like conditional direct branches, indirect branches may lead to more than one
target. Unlike conditional direct branches, which may lead to just two targets,
indirect branches may lead to multiple targets.  Therefore, indirect branch
prediction techniques have traditionally been less efficient than for
conditional direct branch.  Moreover, because of multiple possible targets,
predicting indirect branches needs more resources than predicting conditional
direct branches, leading to more implementation compromises and limitations,
depending on the transistor budget.

In the presence of a large number of conditional cases, it is more efficient to
use a table of targets i.e., a jump table, together with indirect
branches. Having a jump table results in compact code size and reduced branch
depth. On the other hand, it increases the reliance on indirect branch
prediction which has traditionally been less accurate than direct branch
prediction. Moreover, with the increased popularity of interpreted languages and
object oriented languages, many programs rely heavily on jump tables
\cite{kim2007}.

The earliest analysis of splitting the jump table by Bernstein
\cite{Bernstein85} states that the problem of splitting the jump table,
corresponding to case statements in a switch block, into a minimum number of
clusters of a given density is NP complete. Kannan et al. \cite{kannan94} showed
that an $O(n^2)$ can solve this problem as case statements can only have
integral values.

We improve that algorithm by limiting the maximum number of entries in a jump
table to account for the finite number of entries in the indirect branch
predictors. Since an indirect branch can have multiple destinations, most
implementations of indirect branch predictors limit the number of destinations
that it can predict. The indirect branch predictor would perform better when the
size of a jump table is no greater than the number of destinations it can
predict. This modification changes the complexity of the algorithm to $O(nlogn)$
which is bounded by the sorting algorithm used to sort the sequence of case
statements.

\section{Algorithm}
The density of a jump table can be computed as the ratio of number of case
statements (nondefault) to the range of values the case statements cover.  Let
caseitem[i] be the value of i-th case statement. The density $d(i, j)$ of a jump
table having caseitem[i] through caseitem[j], where $i<j$ and caseitem[j] is
included, can be given as \cite{kannan94}:

\[ d(i,j) = \frac{j-i+1}{caseitem[j]-caseitem[i]+1} \]

From the equation, it is clear that the maximum density of a jump table can be
one, i.e., when all the case statements corresponding to each value in the range
is present.

The first step is to sort the array containing the values of the case statements
in ascending order.  This array ($Cases[]$) is the input of the
Algorithm~\ref{algo}. The array $LastElement$ contains the minimum number of
partitions as computed by the algorithm, so the closed range $[i,
  LastElement[i]]$ is a cluster with at most $Max$ case statements, and a
density of at least $D$.  The outer loop of Algorithm~\ref{algo} iterates over
all the case statements, and the inner loop iterates at most $Max$ times: the
complexity of finding the clusters on a sorted array is $O(n)$.

The partitions are then read from the array $LastElement$ and added to
$Partitions$ as closed range of clusters.

\begin{algorithm}[t]
  \DontPrintSemicolon
  \KwData{
    Cases[] = sorted array of case statements values \\
    N = number of elements in Cases \\
    Max = maximum size of jump tables to generate \\
    D = minimum density of a jump table (0 < D <= 1)
  }
  \KwResult{
    Partitions: a list of ranges of indexes in Cases
    identifying partitions satisfying the density and max conditions
  }
  \tcc{MinP contains minimum number of partitions for elements between i and N-1}
  $MinP[N-1]=1$ \\
  \For{i from N-2 to 0} {
    $MinP[i] = 1 + MinP[i+1]$ \\
    $LastElement[i] = i$ \\
    \For{j from min(N-1,i+Max) to i+1} {
      $L = Cases[j] - Cases[i]$ \\
      \If{L $\le$ Max \textbf{and} D < $d(i,j)$} {
        \uIf {j = N-1} {$NumParts = 1$}
        \Else{$NumParts = 1 + MinP[j + 1]$}
        \If {NumParts < MinP[i]} {
          $MinP[i] = NumParts$ \\
          $LastElement[i] = j$
        }
      }
    }
  }
  $i=0$ \\
  \Do{$i < N-1$} {
    $j = LastElement[i]$ \\
    $Partitions.push(i, j)$ \tcp*{closed range [i, j]}
    $i = j+1$
  }
  \caption{Compute minimum number of clusters}
  \label{algo}
\end{algorithm}

\section{Experimental Results}
In general, programs using state machines where a multi-way branch is inside a
loop would benefit the most from this optimization. Experimental results
(Table~\ref{tab:results}) justifies this hypothesis. The SPEC200 benchmarks were
run three times and the median results were considered for the result. We can
see that 253.perlbmk showed noticeable performance improvements while others
remained mostly flat. 253.perlbmk is a Perl interpreter which has a switch
statement inside a loop.  This is exactly the kind of workload that benefits
most from our optimization. 176.gcc also showed some improvement because of the
same reason.

\begin{table}[h!]
\begin{center}
\begin{tabular}{|l|c|}
\hline
SPEC2000 Benchmark & \%Improvement w.r.t baseline \\\hline
164.gzip	&	0\% \\\hline
175.vpr		&	-1\% \\\hline
176.gcc		&	1\% \\\hline
181.mcf		&	-1\% \\\hline
186.crafty	&	0\% \\\hline
197.parser	&	0\% \\\hline
252.eon		&	0\% \\\hline
253.perlbmk	&	10\% \\\hline
254.gap		&	0\% \\\hline
255.vortex	&	0\% \\\hline
256.bzip2	&	0\% \\\hline
300.twolf	&	0\% \\\hline
\end{tabular}
\end{center}
\caption{Runtime Performance improvements as a result of jump-table optimization}
\label{tab:results}
\end{table}


\newpage
\section{Conclusion}
We presented an improved algorithm for clustering case statements by taking into
account the finite size of indirect branch predictors in the
hardware. Experimental results show that this optimization would greatly benefit
programs with heavy use of state machines like interpreters.

\bibliographystyle{abbrv}
{\small
  \bibliography{jt}
}
\end{document}